\documentclass{amsart}
\usepackage{amssymb}
\begin{document}
%
%
%
%
\title{Airy functions\\
in the thermodynamic Bethe ansatz}

\author{Paul Fendley}
\address{
Department of Physics\\
University of Virginia\\
Charlottesville, VA 22901
}

\begin{abstract}
Thermodynamic Bethe ansatz equations are coupled non-linear integral
equations which appear frequently when solving integrable
models. Those associated with models with $N$=2 supersymmetry can be
related to differential equations, among them Painlev\'e III and the
Toda hierarchy. In the simplest such case the massless limit of these
non-linear integral equations can be solved in terms of the Airy
function. This is the only known closed-form solution of
thermodynamic Bethe ansatz equations, outside of free or classical
models.  This turns out to give the spectral determinant of the
Schrodinger equation in a linear potential.

\end{abstract}

\maketitle

A great deal of interesting mathematical physics has arisen from the
study of integrable models of statistical mechanics and field theory.
One interesting area is known as the thermodynamic Bethe ansatz (TBA),
which has proven a useful tool for computing the free energy of an
integrable $1+1$ dimensional system \cite{YY}. One ends up with a set
of coupled non-linear integral equations, the ``TBA equations''.  One
completely-unexpected result was a correspondence between a limit of
these integral equations and some very well-studied non-linear
differential equations, namely the Toda hierarchy \cite{CFIV}. The
purpose of this paper is to extend these results further, and show
that in at least one case there is a closed-form but non-trivial
solution of the integral equations.  Not only is it interesting that
such complicated equations have a simple solution in terms of the Airy
function, but proving it requires some utilizing some very intricate
results involving the Painlev\'e III differential equation
\cite{CFIV,Alyosha,TW}. Moreover, it turns out to be related to the spectral
determinant of the Schrodinger equation in a linear potential
\cite{Voros,Dorey,BLZs}.

The TBA integral equations are generically of the form
\begin{equation}
\epsilon_a(\theta) = m_a \cosh\theta - \sum_b \int {d\theta'\over 2\pi}
\phi_{ab}(\theta-\theta') \ln(1+e^{\mu_b-\epsilon_b(\theta')}).
\label{generic}
\end{equation}
Physically, $T\epsilon_a(\theta)$ is the energy for creating a
particle of type $a$ and rapidity $\theta$ in a thermal bath at
temperature $T$. The $m_a$ are the particle masses over temperature,
while the $\mu_a$ are their chemical potentials over temperature. The
kernels $\phi_{ab}$ are a result of the interactions between
particles.  This and all unlabelled integrals in this paper run from
$-\infty$ to $\infty$.  The free energy per unit length is
\begin{equation}
F= -T^2\sum_a \int {d\theta\over 2\pi}\, m_a \cosh\theta \,
\ln(1+e^{\mu_a-\epsilon_a(\theta)}).
\label{free}
\end{equation}

Here we study TBA equations where the underlying physical system has
$N$=2 supersymmetry. The amazing thing is that solutions of a
particular limit of such TBA equations are simply related to solutions
of a non-linear differential equation \cite{CFIV}.
Particles in an
$N$=2 theory all have a charge, $f_a$, known as their fermion number.
When the chemical potentials are $\mu_a=i\pi
f_a$, a consequence of supersymmetry is that the $\epsilon_a$ in
(\ref{generic}) are all constants and the free energy is $F=0$. This
is known as the Witten index (the usual integer contributions to the
index do not contribute to the free energy per unit length)
\cite{Witten}. The result of \cite{CFIV} is that for chemical
potentials $\mu_a = i(\pi - h)f_a$, one can derive a differential
equation for the order $h$ piece of the free energy. The simplest
cases give the Painlev\'e III differential equation and the Toda
hierarchy.

The TBA equations for the case at hand were derived in \cite{pk}.
They have $\phi_{12}=\phi_{13}=1/\cosh\theta$ with
$\phi_{ab}=\phi_{ba}$ and other $\phi_{ab}=0$, while $m_2=m_3=0$, and
$f_1=0$, $f_2= -f_3= 1$. For small positive $h$, the
functions $A$ and $B$ are defined by $\epsilon_1(\theta)=A(\theta)-
\ln h$, and $\epsilon_2(\theta)=\epsilon_3(\theta)= -hB(\theta)$. The
order $h$ TBA equations are
\begin{eqnarray}
A(\theta)&=& 2u(\theta) - \int {d\theta'\over 2\pi} {1\over
\cosh(\theta-\theta')} \ln (1 + B^2(\theta'))
\label{ABi}\\
B(\theta)&=&\int {d\theta'\over 2\pi} {1\over
\cosh(\theta-\theta')} e^{-A(\theta')}
\label{ABii}
\end{eqnarray}
where $2u(\theta)=m_1\cosh\theta$ here.  In \cite{CFIV}, a physics
proof was given that the resulting free energy is simply related to a
solution of the Painlev\'e III differential equation with variable
$m_1$.  This is a physics proof because one method of computation
gives integral equations, the other the differential equations.  This
result was extended considerably in \cite{Alyosha}. Subsequently, the
equivalence was proven directly and rigorously in \cite{TW}.

A particularly interesting situation is the ``massless'' limit, where
$m_1$ is very small. Then $2u(\theta)=e^\theta$,
because $m_1$ can be removed by redefining $\theta$ by a
shift. The result of this paper is $A$ and $B$ in (\ref{ABi},\ref{ABii})
can be found in closed form
in this massless limit.

\bigskip\noindent
{\bf Result:}\hfill\break
When $u(\theta)=e^\theta/2$, the solution of (\ref{ABi},\ref{ABii}) is
\begin{eqnarray}
\label{resultA}
e^{-A(\theta)} &=& -2\pi {d\over dz} (Ai(z))^2 \\
\label{resultB}
B(\theta) &=& -2\pi\, {d\over dz} \left[Ai
(ze^{i\pi/3}) Ai(ze^{-i\pi/3})\right]
\end{eqnarray}
where $z=(3e^\theta/4)^{2/3}$ and $Ai(z)$ is the Airy function.

\bigskip\noindent To check the normalization, note that
(\ref{ABi},\ref{ABii}) imply that $e^{-A(\theta)} \to 2/\sqrt{3}$ and
$B(\theta)\to 1/\sqrt{3}$ as $\theta\to -\infty$, in agreement with
the limits of the appropriate Airy functions as $z\to 0$.  $Ai(z)$ is
a solution of the differential equation $w^{\prime\prime}=zw$, while
$e^{-A(\theta)}$ and $B(\theta)$ solve $w^{\prime\prime\prime} - 4zw'=6w$.

As far as I can tell, it is difficult to prove the result by direct
substitution into the integrals, but requires utilizing some
additional structure. First, define the integral operator $K$ which
maps functions to functions with the kernel
$$K(\theta,\theta')= 2{E(\theta)E(\theta') \over e^\theta +
e^{\theta'}},$$ where $$E(\theta)=e^{\theta/2} e^{-u(\theta)}.$$
Solutions of Painlev\'e III can be expressed in terms of this operator
$K$ \cite{MTW,CFIV,Alyosha} when $2u=m_1\cosh\theta$.  The functions
$Z_+$ and $Z_-$ are defined as
\begin{equation}
Z_+=e^{-2\theta/3} (I+K)^{-1}E
\qquad Z_-=e^{-\theta/3}(I-K)^{-1}E
\label{zdef}
\end{equation}
These $Z_\pm$ are simply related to the $Q\mp P =(I\pm K)^{-1}E$ used
in \cite{TW}.  It was proven in \cite{CFIV,Alyosha,TW} that functions
$A$ and $B$ solve (\ref{ABi},\ref{ABii}) if
\begin{eqnarray}
\label{TWi}
e^{-A(\theta)}&=& 4\pi Z_+(\theta)Z_-(\theta)\\
B(\theta)&=&4\pi e^{i\pi/3} e^{u(\theta+i\pi/2)+u(\theta-i\pi/2)}
Z_+(\theta-i\pi/2)Z_-(\theta+i\pi/2) -i
\label{TWii}
\end{eqnarray}
The functions (\ref{TWi},\ref{TWii}) obey (\ref{ABi},\ref{ABii}) for
any entire $u(\theta)$ obeying $u(\theta)=u(\theta+2\pi i)$
\cite{Alyosha,TW}, not only the special case $u(\theta)=e^\theta/2$
analyzed in this paper.

With the similarity between (\ref{ABi},\ref{ABii}) and
(\ref{TWi},\ref{TWii}), proving the result is equivalent to proving
that 
\begin{equation}
Z_+(\theta)=Ai(z) \qquad Z_-(\theta) = -Ai\,^\prime(z)
\label{Zairy}
\end{equation}
when $u(\theta)=e^\theta /2$. The expression (\ref{resultB}) for
$B(\theta)$ follows by using a few Airy-function identities.
The expressions for $Z_\pm$ in (\ref{Zairy}) can be proven by directly
evaluating the integral in their definition. Specifically,
(\ref{zdef}) requires that
\begin{equation}
E(\theta)= e^{2\theta/3} Z_+(\theta) + 2E(\theta) \int d\theta'
Z_+(\theta') e^{-e^{\theta'}/2}\,
{e^{7\theta'/6} \over e^\theta + e^{\theta'}}
\label{integ}
\end{equation}
Defining $x\equiv e^\theta/2$ and using the expression of an Airy
function in terms of a Bessel function, the integral is proportional to
$$\int_0^\infty
dx\ e^{-x} {\sqrt{2x} \over e^\theta + 2x} K_{1/3}(x).$$ This
can be looked up in \cite{Erdelyi}, or evaluated by using Mellin
transforms, under which the Bessel function $K_\nu$ has nice
properties.  One indeed finds that (\ref{integ}) and the analogous
relation for $Z_-$ are true when $Z_\pm$ are given by
(\ref{Zairy}). Since the solution with the appropriate analyticity
properties (no zeroes and bounded in the strip $|Im \theta|<\pi/2$) is
unique \cite{TW}, the result is proven.

The functions $Z_+(\theta)$ and $Z_-(\theta)$ are interesting in their
own right. They arose as a sort of partition function of the
boundary-sine Gordon problem \cite{FLS} at coupling $g$=2/3, and
equivalently as the continuum version of the Baxter $Q$-operator
\cite{Baxter} in studies of conformal field theory \cite{BLZ}.  The
$Q$ operator gives the generating function of the conserved quantities
(local and non-local) of the theory.  Up to normalization, the
functions $Z_\pm$ are $Z_{BSG}(z,\pm 1/3)$ in \cite{FLS}, and $\langle
Q(z)\rangle$ at $p=\pm 1/6$ in \cite{BLZ}.  Thus the result here
provides the only case (other than where the system is free or
classical) where these quantities can be computed explicitly. In fact,
it provides strong evidence that the results of \cite{FLS,BLZ} are
applicable in the repulsive regime $g>1/2$ as conjectured. For
example, it shows that the zeroes of the eigenvalues of the
$Q$-operator obey the pattern conjectured in \cite{BLZ}. The ``quantum
Wronskian'' of \cite{BLZ} becomes equivalent to the Proposition of
\cite{TW}, and both are easily verified by using the (ordinary)
Wronskian of the Airy functions.  Also, by using the series expansion
of \cite{FLS}, it gives a strange sequence of identities of sums of
products of gamma functions.

Recently, the work of \cite{BLZ} also arose in a completely new
context. Namely, it was observed in \cite{Dorey} that the spectral
determinant for the Schrodinger equation in a potential $|x|^{\alpha}$ for
any $\alpha$ is given precisely by the vacuum eigenvalue of this
Baxter $Q$-operator. A non-zero angular momentum can be added, and
this correspondence still holds \cite{BLZs}. This is because the
quantum Wronskian is equivalent to a set of functional relations for
the spectral determinant derived in \cite{Voros,BLZs}. The result above
then means that Airy function is the spectral determinant for the
Schrodinger equation in a linear potential, a result shown directly in
\cite{Voros}. It would be most interesting to understand how to extend
this result.

Since this system discussed in this paper provides the simplest
example of the differential equation/TBA correspondence \cite{CFIV},
it seems likely that there is a simple solution of the massless limit
of any TBA equations of this type. What is not yet known outside of
this case is the detailed analysis of the differential equation of
\cite{MTW}, which was vital to the results of \cite{TW}. Given how
beautifully the Airy function solved the problem here, it would be
quite interesting to see how this result is generalized.

\bigskip\bigskip 

This work was mostly done in spring 1996, while I was a postdoc at the
University of Southern California.  I thank Hubert Saleur and the
Packard Foundation for support then.  It was presented in 1997 at the
Supersymmetry and Integrable Models Workshop, University of
Illinois-Chicago, and the International Workshop on Statistical
Mechanics and Integrable Systems in Coolongatta. I thank the
organizers for their hospitality. I thank P. Dorey for conversations,
and S. Lukyanov for urging me to write this up despite the late
date. Currently, my work is supported by a DOE OJI Award, a Sloan
Foundation Fellowship, and by NSF grant DMR-9802813.

\bigskip\bigskip

%
\renewcommand{\baselinestretch}{1}

\end{document}